\documentclass[twocolumn]{aastex631}

\usepackage{graphicx}
\usepackage{amsmath}
\usepackage{xcolor}

\shorttitle{The Pop III.1 Cosmic Infrared Background}
\shortauthors{Tataroglu et al.}

\begin{document}

%\title{Cosmic Background Signatures of Flash Ionization by Population III.1 Supermassive Stars}

\title{The Cosmic Infrared Background from Population III.1 Supermassive Stars and Black Holes}

\author{Emir Tataroglu}
\affiliation{Department of Physics, University of Chicago, Chicago, IL 60637, USA}
\affiliation{Department of Astronomy, University of Virginia, Charlottesville, VA 22904, USA}
%\altaffiliation{\raggedright Present address: Department of Physics, University of Chicago, Chicago, IL, USA}

\author{Jonathan C. Tan}
\affiliation{Department of Astronomy, University of Virginia, Charlottesville, VA 22904, USA}
\affiliation{Virginia Institute of Theoretical Astronomy, University of Virginia, Charlottesville, VA 22904, USA}
\affiliation{Department of Physics \& Astronomy, Chalmers University of Technology, Gothenburg, Sweden}
\affiliation{Dept. of Astronomy \& Astrophysics, UC San Diego, San Diego, CA 92093, USA}

\author{Rosalia D. O'Brien}
\affiliation{Department of Astronomy, University of Maryland, College Park, MD 20742, USA}
\affiliation{Astrophysics Science Division, Code 660, NASA Goddard Space Flight Center, 8800 Greenbelt Rd., Greenbelt, MD 20771, USA}
\affiliation{Center for Research and Exploration in Space Science and Technology, NASA/GSFC, Greenbelt, MD 20771 USA}

\author{Devesh Nandal}
\affiliation{Department of Astronomy, University of Virginia, Charlottesville, VA 22904, USA}
\affiliation{Center for Astrophysics, Harvard \& Smithsonian, 60 Garden Street, Cambridge, MA 02138, USA}

\correspondingauthor{Emir Tataroglu}
\email{[emirt@uchicago.edu]}

\begin{abstract}
We investigate the cosmic infrared background (CIB) that is expected to arise from the Population (Pop) III.1 supermassive black hole (SMBH) formation scenario, which involves early SMBH seeding via supermassive primordial star formation enabled by dark matter annihilation heating. We model the emission from Pop III.1 supermassive stars, including their growth phase, flash ionization phase, and early black hole accretion phase, as well as accompanying emission from lower-mass Pop III.2 stars. The CIB is composed of stellar and SMBH accretion disk continuum radiation, nebular continuum emission, and Ly$\alpha$ radiation redistributed through scattering in the neutral intergalactic medium. 
Based on the feedback regulated Pop III.1 model of Petkova et al. (2026), our fiducial model assumes $\sim 10^5\,M_\odot$ Pop~III.1 stars forming by $z_{\rm flash}=20$ with a comoving number density of $\sim0.2\,{\rm cMpc}^{-3}$. For this model, the combined stellar contribution is less than 1\% of existing direct and diffuse CIB measurements. However, the contribution to the CIB from $z\sim6-20$ SMBH accretion is expected to be larger than the Pop III.1 stellar component for average Eddington ratios $\lambda_{\text{Edd}} \sim 0.01$. Latest CIB observations place constraints of $\lambda_{\text{Edd}} \lesssim 0.3$ for the fiducial Pop III.1 model during these epochs.
%and could match amplitudes of the potential observed CIB excess for $. 
%We discuss the distinctive spectral signatures of the Pop III.1 CIB model and  
%The Population III.1 scenario can thus explain the bulk of the the CIB excess given AGN Eddington ratios above $\lambda_{\text{Edd}} \sim 0.5 $ for otherwise fiducial model parameters. 
%Even in less luminous parameter spaces, it produces distinctive spectral features that may provide an independent observational test of primordial supermassive stars and early black-hole seed formation.
\end{abstract}

\keywords{Population III stars --- Extragalactic background light --- Cosmology --- Early universe --- Reionization}

\section{Introduction}

The extragalactic background light (EBL) is the accumulated emission from all cosmic sources over time, and its amplitude and spectrum are significant constraints on cosmological models. However, since the first absolute near-infrared measurements with COBE/DIRBE and IRTS, direct estimates of the EBL in the infrared regime, i.e., the Cosmic Infrared Background (CIB), have often exceeded the integrated galaxy light inferred from resolved galaxy counts, suggesting that an unresolved or diffuse component exists \citep{Hauser1998DIRBE,DwekArendt1998,Gorjian2000,Cambresy2001,Matsumoto2005IRTS}. Many possible explanations have been put forth to resolve this, including: residual foreground contamination from zodiacal light \citep{Kelsall1998DIRBEZodi}; faint or low-surface-brightness galaxy emission missed by galaxy counts \citep{Driver2016EBLCounts}; diffuse intra-halo light from tidally stripped stars \citep{Cooray2012IHL,Zemcov2014CIBER}; and high-redshift sources such as Population (Pop) III stars \citep{Santos2002FirstStarsCIB,salvaterra2003}.

It is easy to see why Pop III stars, which are defined to be ``metal-free'' (i.e., composition set by Big Bang nucleosynthesis), are natural candidates for the CIB discrepancy. At high redshift, their stellar spectra are redshifted into the near-infrared, while interactions with the surrounding primordial gas and neutral intergalactic medium (IGM) reshape the emitted spectrum by reprocessing ionizing photons into recombination lines, nebular free-free, free-bound, and two-photon continuum emission, and resonant Ly-$\alpha$ scattering in the neutral IGM. Depending on the physical assumptions made, the observed radiation from Pop III has been shown to be able to reach the magnitudes of the CIB discrepancy \citep{Schaerer2002,Santos2002FirstStarsCIB,salvaterra2003}. This possibility has gained renewed interest with JWST, which has now reported candidate Pop III signatures in high-redshift systems, including possible strong He II emission without associated metal lines and ultrafaint metal-poor galaxy candidates \citep{Maiolino2024GNz11PopIII,Fujimoto2025GLIMPSE}.

The Pop III.1 model for supermassive black hole (SMBH) formation \citep{Banik2019PopIII1Seeds,Singh2023PopIII1SeedsII,Cammelli2025PopIII1SeedsIII,Petkova2026PopIIISeedsIV} \citep[see][for a review]{Tan2024OriginPopIII1Seeds} proposes that the very first Pop III stars to form in each local region of the Universe, i.e., ``Pop III.1'', are thus unaffected by feedback from other astrophysical sources leading to pristine initial conditions \citep{McKeeTan2008} and can then, under the influence of WIMP dark matter particle annihilation \citep{2008PhRvL.100e1101S,2009ApJ...692..574N,2015ApJ...799..210R,2025arXiv250700870N,2025arXiv251000216T}, grow to become supermassive, i.e., $\sim10^5\:M_\odot$, stars that are the seeds for all SMBHs.

%A particularly promising  Pop III model is the Population III.1 supermassive black hole (SMBH) formation theory, in which rare supermassive primordial stars form in pristine early halos of mass $\sim 10^6 M_{\odot} $ and later collapse into massive black-hole seeds (\cite{Tan2025FlashIonization}, \cite{Cammelli2025PopIII1SeedsIII}, \cite{Banik2019PopIII1Seeds}, \cite{McKeeTan2008}). 

The fiducial Pop III.1 model involves the supermassive stars being sources of strong H-ionizing luminosity. They drive rapidly expanding HII regions and produce a short-lived flash-ionization of the early Universe at $z \sim 20-30$ \citep{Tan2025FlashIonization}, with this process self-regulating the abundance of SMBHs \citep{Petkova2026PopIIISeedsIV}. This scenario involving an early phase of cosmic ionization and an increased CMB scattering optical depth has the potential to alleviate several current cosmological tensions, including discrepant CMB-based measurements of $H_0$, inference of negative neutrino masses, and dynamical dark energy 
%preferences, through its additional contribution to the CMB scattering optical depth 
\citep{2025JCAP...08..082A,2025PhRvD.112d3541J,Tan2025FlashIonization,2026PhRvD.113h3515A,2026PhRvL.136h1002S,2026ApJ..1005L..32T,2026arXiv260619459A}.

%the resulting massive black-hole seeds. The ionized gas then recombines back toward a largely neutral state on a $\sim 10\,{\rm Myr}$ timescale. 

In this paper we present a first calculation of the Pop~III.1 contribution to the CIB. We calculate the emission from Pop III.1 supermassive stars in both their protostellar (growth) phase and in their flash ionization phase. We also include contributions from lower-mass Pop III.2 stars, i.e., metal-free stars that have been impacted by external feedback \citep[][]{2006MNRAS.373..128G,McKeeTan2008}.
%and their primary flash-ionization and recombination signal, we also consider two associated contributions: emission from Pop III.2 stars, i.e., lower-mass metal-free stars that form from primordial gas whose thermal and chemical state has been altered by earlier radiative feedback, 
Finally, we also include an estimate for the contribution to the CIB from early ($z\gtrsim6$) accretion onto Pop III.1-seeded SMBHs.
%Due to the significantly higher mass and lower abundance of Pop III.1 sources, their potential contribution to the EBL can be very different than traditional lower-mass Pop III models. This motivates a direct calculation of the EBL imprint of the Pop III.1 flash-ionization scenario. 
%We model the resulting signal from Pop III.1 and Pop III.2 sources as well as black hole accretion, 
In these models we consider direct stellar/SMBH emission, nebular continuum emission, and Ly$\alpha$ radiation redistributed by scattering in the neutral IGM. We focus on the near and mid-infrared  frequencies, comparing the predicted intensity with direct EBL and diffuse light measurements. We are then able to assess whether the Pop III.1 model can account for a significant fraction of the CIB.

\section{Pop III.1 CIB Model}

Given a medium of emissivity $\epsilon_\nu(\nu,z)$ and effective optical depth $\tau_{\rm eff}$ in a spatially flat FLRW metric, the observed present-day specific intensity is given by the standard path integral 
\begin{equation}
I_{\nu_0} =
{\frac{1}{4\pi}}
\int
\epsilon_\nu(\nu,z)
e^{-\tau_{\rm eff}}
{dl \over dz}\,dz ,
\end{equation}
where $\nu = \nu_0 (1+z)$ is the emitted-frame frequency corresponding to the observed frequency,  $\nu_0$, 
\begin{equation}
{dl \over dz}={c \over H_0(1+z)E(z)}
\end{equation}
is the proper line element per unit redshift, and 
\begin{equation}
E(z)=\left[\Omega_M(1+z)^3+\Omega_\Lambda\right]^{1/2}
\end{equation}
is the dimensionless Hubble parameter for a spatially flat $\Lambda$CDM cosmology. The integral is evaluated over the redshift interval during which the source population is present. We neglect absorption outside of the Ly\(\alpha\) redistribution included below, so that $e^{-\tau_{\rm eff}}=1.$

\subsection{Source Emissivity}

Following \citet{salvaterra2003}, we define a comoving emissivity 
\begin{equation}
\epsilon_\nu(\nu,z)=l_\nu(\nu,z)\rho_\star(z),
\end{equation}
where $l_\nu(\nu,z)$ is the specific luminosity of the population and $\rho_\star(z)$ is the stellar mass density per unit comoving volume.
We can assign each dark matter halo of mass $M_h$ a corresponding stellar mass,
\begin{equation}
M_\star=f_\star{\Omega_b \over \Omega_M}M_h ,
\end{equation}
leading to a stellar mass density per comoving volume,
\begin{equation}
\rho_\star(z)=
{\Omega_b \over \Omega_M}
\int_{M_{\rm min}(z)}^\infty
f_\star(M_h,z) n(M_h,z)M_h\,dM_h .
\end{equation}

Here: \(n(M_h,z)\) is the halo mass function \citep[for which we adopt the analytic model of][]{press1974}; $f_\star(M_h,z)$ is the fraction of baryons in a halo that are able to form stars, i.e., the star formation efficiency, which can vary with halo mass, redshift and other particular properties pertaining to the Pop III.1 model; and \(M_{\rm min}(z)\) is the minimum halo mass allowed to form stars. 
To account for finite lifetime of sources, especially lifetimes of Pop III stars, we modify the \cite{salvaterra2003} model and write the emitting stellar mass density as
\begin{equation}
\rho_{\star}(z)
=
\rho_{\star,0}(z)\,\mathcal{W}_\star(z),
\end{equation}
where we introduce a survival factor \(\mathcal{W}_\star\). In the simplest case,
\begin{equation}\label{eq:W}
\mathcal{W}_\star(z)=
\Theta\!\left[t_\star-\left(t(z)-t_{\rm form}\right)\right],
\end{equation}
where $\Theta$ is the Heaviside step function, so that a stellar population contributes while its age is less than \(t_\star + t_{\rm form}\), and is removed after that lifetime. Note, the formation time, $t_{\rm form}$, is introduced to allow for a distinct protostellar phase of Pop III.1 supermassive stars.

%\textcolor{red}{(This is the only modification in this subsection to the SF model)}.

In the general case of a stellar population, its specific luminosity is
\begin{equation}
l_\nu(\nu,z)=
\int_{m_{\star,l}}^{m_{\star,u}}
F(\nu,m_\star,z)\phi(m_\star)\,dm_\star ,
\end{equation}
where the $\phi(m_\star)$ is the chosen initial mass function (IMF), $m_{\star,l}$ and $m_{\star,u}$ are the upper and lower mass bounds of the IMF, respectively, and $F(\nu,m_\star,z)$ is the stellar spectral energy distribution (SED). The IMF is normalized so that
\begin{equation}
\int_{m_{\star,l}}^{m_{\star,u}}\phi(m_\star)\,dm_\star=1.
\end{equation}

The SED is decomposed into three terms: 
\begin{equation}
F(\nu,m_\star,z)
=
l_\nu^{\rm star}
+
l_\nu^{\rm neb}
+
l_\nu^{\rm Ly\alpha}.
\end{equation}
These terms correspond to direct stellar (or SMBH) continuum emission, nebular continuum emission, and Ly\(\alpha\) photons redistributed by scattering in the surrounding neutral IGM, respectively. 
The nebular continuum luminosity is given by
\begin{equation}
l_\nu^{\rm neb}
=
{\gamma_{\rm tot} \over \alpha_B}
(1-f_{\rm esc})q({\rm H}),
\end{equation}
where \(\alpha_B\) is the Case B recombination coefficient, \(q({\rm H})\) is the ionizing photon rate per stellar mass, and
\begin{equation}
\gamma_{\rm tot}
=
\gamma_{\rm H I}
+
\gamma_{2q}
+
\gamma_{\rm He I}{n({\rm He}^+) \over n({\rm H}^+)}
+
\gamma_{\rm He II}{n({\rm He}^{++}) \over n({\rm H}^+)} 
\end{equation}
is the total continuous emission coefficient, which includes hydrogen free-free and free-bound emission, the hydrogen two-photon continuum, and helium free-free and free-bound emission.
The ionizing photon production rate per unit stellar mass is given by
\begin{equation}
q({\rm H})
=
\int_{\nu_{\rm LL}}^\infty
{l_\nu^{\rm star}\over h\nu}\,d\nu .
\end{equation}
A fraction, \((1-f_{\rm esc})\), of these photons is absorbed locally and causes recombination emission. The resulting Ly\(\alpha\) luminosity, which is separate from the nebular emission of $l_\nu^{\rm neb}$, is then redistributed by resonant scattering in the neutral IGM:
\begin{equation} 
l_\nu^{\rm Ly\alpha}
=
c_{\rm Ly\alpha}(1-f_{\rm esc})q({\rm H})
\Phi(\nu_{\rm Ly\alpha}-\nu,z).
\end{equation}
where $c_{\mathrm{Ly}\alpha} = 1.04 \times 10^{-11}\,\mathrm{erg}$ \citep{Schaerer2002}, and 
\begin{equation} \label{eq:scattering}
\Phi(\Delta\nu,z)
=
\nu_\star(z)\Delta\nu^{-2}
\exp\left[-{\nu_\star(z)\over \Delta\nu}\right],
\qquad \Delta\nu>0 ,
\end{equation}
is the Ly\(\alpha\) scattered line profile for a homogeneous, expanding IGM \citep{Santos2002FirstStarsCIB}, with
\begin{equation}
\nu_\star(z)
=
1.5\times10^{12} 
\left({\Omega_b h^2 \over 0.019}\right)
\left({h \over 0.7}\right)^{-1}
{(1+z)^3 \over E(z)}
\ {\rm Hz}.
\end{equation}

For the helium contribution to the nebular continuum, we adopt a primordial helium mass fraction \(Y_{\rm He}=0.244\), corresponding to \(n_{\rm He}/n_{\rm H}=Y_{\rm He}/[4(1-Y_{\rm He})]\). The relative He\(^{+}\) and He\(^{++}\) abundances are estimated from the hardness of the adopted blackbody spectrum and are used only to set the helium free--free and free--bound continuum coefficients.

%For the Larson IMF (see Figure \ref{fig:IMF} and \cite{salvaterra2003}),
%\begin{equation}
%\phi(M)\propto
%M^{-1}
%\left(1+{M\over M_c}\right)^{-1.35}.
%\end{equation}

%\begin{figure} 
    %\centering
    %\includegraphics[width=1 \linewidth]{S&FIMF.png}
   %\caption{Observed intensity from Population III stars for different low mass IMFs. We consider the redshift range $z = 20$ to $z_{end} = 8.8$, assuming sources exist for the entire duration. The halo mass function  from \cite{press1974} is adopted, with a fixed star formation efficiency $f_\star = 0.21$ and no overdensity: $\delta_{IGM} = 1$. This leads to stellar number densities multiple orders of magnitude above those assumed for the high mass Pop III.1 scenario. \textcolor{red}{This graph is not novel, and so it is temporary}}
   %\label{fig:IMF}
%\end{figure}

\subsection{Pop III.1 Sources}

For Pop III.1 sources we keep the same spectral decomposition as described above, but model the sources as finite lifetime supermassive stars that all form at a common redshift reaching their post-formation, ionizing phase at a fiducial $z_{\rm flash}=20$ (and corresponding time $t_{\rm flash}=t_0=180\:$Myr). We assume each star accretes to a final mass of $m_{\rm \star,III.1} = 10^5 M_\odot$, with a fiducial formation timescale of 30~Myr, i.e., an average accretion rate of $3\times10^{-3}\:M_\odot\:{\rm yr}^{-1}$ \citep{2010AIPC.1294...34T}. We adopt a fiducial post-formation lifetime of 10~Myr, i.e., somewhat lengthened compared to standard stellar lifetimes by residual WIMP annihilation heating. In both the formation and post-formation phases we approximate Pop~III.1 stars as having bolometric luminosities equal to the Eddington luminosity, i.e.,
\begin{equation}
    L_{\rm bol,III.1} = 3.2\times10^9 \left(\frac{m_{\rm *,III.1}}{10^5\:M_\odot}\right) L_\odot.
\end{equation}
We assume Pop III.1 stellar photospheres are black bodies. In the formation phase we adopt $T=6,000\:$K \citep{2025arXiv250700870N}, while in the post-formation phase we assume $T = 10^5\:$K.  

The comoving emissivity from the Pop III.1 source population is then calculated as described above, i.e.,
\begin{equation}
\epsilon_{\rm \nu,III.1}(\nu,z)
=
l_{\rm \nu,III.1}(\nu,z)\rho_{\rm \star,III.1}(z).
\end{equation}
To consider stellar lifetime, we write the active Pop III.1 stellar mass density as
\begin{equation}
\rho_{\rm \star,III.1}(z)
=
\rho_{\rm \star,0,III.1}\,
\mathcal{W}_{\rm III.1}(z),
\end{equation}
where \(\mathcal{W}_{\rm III.1}\) is the Pop III.1 survival factor (see eq.~\ref{eq:W}). 
%For a Pop III.1 flash at \(t_0=t(z_{\rm flash})\), for the post-formation phase we adopt
%\begin{equation}
%\mathcal{W}_{\rm III.1}(z)
%=
%\Theta\!\left[t(z)-t_0\right]
%\Theta\!\left[t_\star-\left(t(z)-t_0\right)\right].
%\end{equation}
This turns the main ionizing Pop III.1 contribution on at \(z_{\rm flash}\) and removes it once the source age exceeds \(t_\star\). Similarly, the formation phase is assumed to extend for the prior period of 30~Myr before $t_0$.

We adopt a fiducial Pop~III.1 source number density of $
n_{\rm III.1} = 0.2\: {\rm Mpc}^{-3}$ and thus a flash stellar mass density of 
\begin{equation}
\rho_{\rm \star,III.1}
=
n_{\rm III.1} m_{\rm \star,III.1}.
\end{equation}
This fiducial source abundance should be regarded as parameter of the model \citep{Banik2019PopIII1Seeds,Singh2023PopIII1SeedsII,Cammelli2025PopIII1SeedsIII}. However, self-consistent feedback regulated models find typical values at about this level \citep{Tan2024OriginPopIII1Seeds,Sanati2025PopIII1Feedback,Petkova2026PopIIISeedsIV}.

Writing the same normalization in terms of a halo star-formation efficiency gives
\begin{equation}
\rho_{\rm \star,III.1}
=
f_{\rm \star,III.1}\frac{\Omega_b}{\Omega_M}
\mathcal{M}_h(z_{\rm flash}),
\end{equation}
where we define the halo mass integral
\begin{equation}
\mathcal{M}_h(z_{\rm flash})
=
\int_{M_{\rm min}(z_{\rm flash})}^\infty
n(M_h,z_{\rm flash})M_h\,dM_h .
\end{equation}
We thus adopt a redshift-dependent star formation efficiency in order to enforce the desired number density of sources in the evolving halo population due to the \cite{press1974} formalism.  

%Therefore, the star formation efficiency that gives the desired number density of sources is 

%\begin{equation}
%f_{\rm \star,III.1}
%= {\Omega_M \over\Omega_b }
%{n_0 m_{\rm *f,III.1} 
%\over{ }
%\mathcal{M}_h(z)}.
%\end{equation}

%Note, the nebular luminosity term in the stellar spectrum is unchanged from the standard Pop III model, and the Ly\(\alpha\) term is only affected by the modified $q(H)$ coming from the Pop III.1 sources. For the direct stellar term, we include only the continuum photons that are not assigned to the recombination and scattered Ly\(\alpha\) channels, so that we avoid double-counting photons (see Figure \ref{fig:flashdecomposition}). 

%As an optimistic approximation for the growth phase of the Pop III.1 sources, we consider a $100 $ ${\rm Myr}$ period of the same comoving number density as the flash phase and reduced temperature $T = 6000 K$, occurring precisely between the flash and  $100 $ ${\rm Myr}$ prior (see Figure \ref{fig:flashdecomposition}). Since the growth-phase temperature is too low to produce a significant ionizing photon rate, we include it only as direct blackbody emission.
 
\subsection{Pop III.2 Sources}

We include Pop III.2 component associated with lower-mass metal-free stars forming in minihalos affected by earlier radiative feedback. In contrast to the Pop III.1 channel, where the abundance is imposed by the SMBH-seed number density, we normalize Pop III.2 by assigning one star or smaller stellar population of characteristic mass $m_{\rm \star,III.2}$ to each eligible minihalo. Given that $n_{\rm III.1}\ll n_{\rm III.2}$, the comoving source density is approximately described by
\begin{equation}
n_{\rm III.2}(z)
=
\int_{M_{\rm min}}^\infty n(M_h,z)\,dM_h ,
\end{equation}
where we adopt a minihalo threshold mass of $M_{\rm min}=10^6\,M_\odot$. The corresponding active stellar mass density is
\begin{equation}
\rho_{\rm \star,III.2}(z)
=
m_{\rm \star,III.2}\,
n_{\rm III.2}(z)\,
\mathcal{W}_{\rm III.2}(z),
\end{equation}
where the survival factor is treated analogously to the Pop III.1 flash model. However, here we will also consider some cases in which the source is treated as a stellar population that has a relatively long duration for its existence.

We use the same radiative channels as for Pop III.1: direct blackbody stellar emission, nebular continuum, and Ly\(\alpha\) radiation redistributed by the neutral IGM. However, Pop III.2 differs in its normalization: its abundance is set by the minihalo population rather than by an imposed SMBH-seed density. This allows the Pop~III.2 component to test whether the much more numerous, lower-mass primordial stellar population can produce a background comparable to the rarer but far more luminous Pop III.1 sources.
%(see Figures \ref{fig:poptotal} and \ref{fig:popiisweep}). 

\subsection{AGN Contribution from Pop III.1 SMBH Seeds}\label{section:AGN}

We also consider a high-redshift AGN contribution from the black-hole remnants of the Pop III.1 sources. We assume that all Pop III.1 sources collapse into  SMBHs starting $t_{\rm \star,III.1}$ years after their formation epoch and continue accreting until $z_{\rm end}$, maintaining a constant comoving number density, $n_{\rm III.1}$. The choice of the fiducial value of $z_{\rm end}=6$ is based on the high redshift limit of the \citet{andrews2018csed} model for the contribution to the EBL of ``normal'' galaxies and AGN.
During this high redshift AGN phase, each SMBH is assumed to accrete continuously at a rate that is a fixed fraction of its own growing Eddington accretion rate. Assuming a radiative efficiency $\epsilon=0.1$, we define
\begin{equation}
  \dot M_{\rm Edd}(M_{\rm BH}) \equiv \frac{L_{\rm Edd}(M_{\rm BH})}{\epsilon c^2}
\end{equation}
where the Eddington luminosity is 
\begin{equation}
     L_{\rm Edd}(M)=1.26\times10^{38}\left(\frac{M}{M_\odot}\right){\rm erg\,s^{-1}},
\end{equation}
and parameterize the accretion by the Eddington ratio
\begin{equation}
  \lambda_{\rm Edd} \equiv \frac{\dot M_{\rm BH}(t)}{\dot M_{\rm Edd}[M_{\rm BH}(t)]},
\end{equation}
so that the total mass accretion rate to the black hole system is
\begin{equation}
  \dot M_{\rm tot}(t) = \lambda_{\rm Edd}\,\dot M_{\rm Edd}[M_{\rm BH}(t)] = \frac{\lambda_{\rm Edd}}{\epsilon c^2}\,L_{\rm Edd}[M_{\rm BH}(t)].
\end{equation}
However, a fraction $\epsilon$ of the inflowing rest mass is radiated, so the black hole grows at a rate
\begin{equation}
\dot M_{\rm BH}=(1-\epsilon)\dot M_{\rm tot}\propto M_{\rm BH} 
\end{equation}
and, at fixed Eddington ratio, its mass grows exponentially,
\begin{equation}\label{eq:SMBHgrowth}
  M_{\rm BH}(t) = M_{\rm BH,0}\,\exp\!\left(\frac{t-t_0}{t_{\rm acc}}\right),
\end{equation}
where the accretion time is given by
\begin{equation}
    t_{\rm acc} = \frac{\epsilon}{1-\epsilon}\,\frac{c^2 M_\odot}{\lambda_{\rm Edd}\,L_{\rm Edd,\odot}}
\end{equation}
and where $t_0$ marks the start of the AGN phase.

Over the AGN duration $\Delta t_{\rm AGN}$ we define the total growth factor
\begin{equation}
f_{\rm grow} \equiv \frac{M_{\rm BH}(z_{\text{end}})}{M_{\rm seed}} = \exp\!\left(\frac{\Delta t_{\rm AGN}}{t_{\rm acc}}\right) 
\end{equation}
and the accreted mass and total radiated energy are thus 
\begin{equation}
    \Delta M_{\rm BH}=(f_{\rm grow}-1)M_{\rm seed}, \qquad E_{\rm AGN}=\frac{\epsilon}{1-\epsilon}\,\Delta M_{\rm BH}\,c^2.
\end{equation}

Following the treatment of quasar disks in \citet{GoodmanTan2004}, we model the instantaneous accretion spectrum as a sum of blackbody annuli from an optically thick disk. Each annulus of the accretion disk at effective temperature $T(R)$ has specific intensity given by the Planck function,
\begin{equation}
  B_\nu(T) = \frac{2h\nu^{3}}{c^{2}}\,\frac{1}{\exp\!\left(h\nu/k_{\rm B}T\right)-1}.
\end{equation}
By integrating over radius, we get the disk luminosity per unit frequency
\begin{equation}
L_\nu^{\rm AGN}(t) = 4\pi^2\int_{R_{\rm in}}^{R_{\rm out}} B_\nu\!\left[T(R;M_{\rm BH}(t),\dot M(t))\right] R\,dR ,
\end{equation}
with the \cite{Pringle1981} thin-disk temperature profile
\begin{equation}
T(R;M_{\rm BH},\dot M) = \left[\frac{3GM_{\rm BH}\dot M}{8\pi\sigma_{\rm SB}R^3}\left(1-\sqrt{R_{\rm in}/R}\right)\right]^{1/4} ,
\end{equation}
and inner and outer radii $R_{\rm in}=3R_S$, $R_{\rm out}=1000R_S$, where $R_S=2GM_{\rm BH}/c^2$ is the instantaneous Schwarzschild radius. 

The resulting AGN spectrum is passed through the same nebular continuum and redistributed Ly$\alpha$ channels used for the Pop III.1 stellar emission, using the same Lyman-limit threshold, escape fraction $f_{\rm esc}=0$, and IGM scattering profile. 

%Because $M_{\rm BH}(t)$ and $\dot M(t)$ evolve together self-consistently we discretize the growth history into $N$ steps spaced uniformly in $\log M_{\rm BH}$ between $M_{\rm seed}$ and $M_{\rm BH}(z=8)$ evaluate $L_\nu^{\rm AGN}$ with the appropriate $\dot M$ at each mass step, and average to obtain a time-averaged spectrum. The ratio of the raw disk bolometric luminosity to $\epsilon\dot M c^2$ depends only on the fixed disk geometry ($R_{\rm in}/R_{\rm out}$), not on $M_{\rm BH}$ or $\dot M$ individually, so we rescale the time-averaged spectrum by this single constant factor to enforce $L_{\rm bol}^{\rm AGN}=\epsilon\dot M c^2$ exactly at every step.

\section{Results}

\begin{figure*}
\centering
\includegraphics[width=1\linewidth]{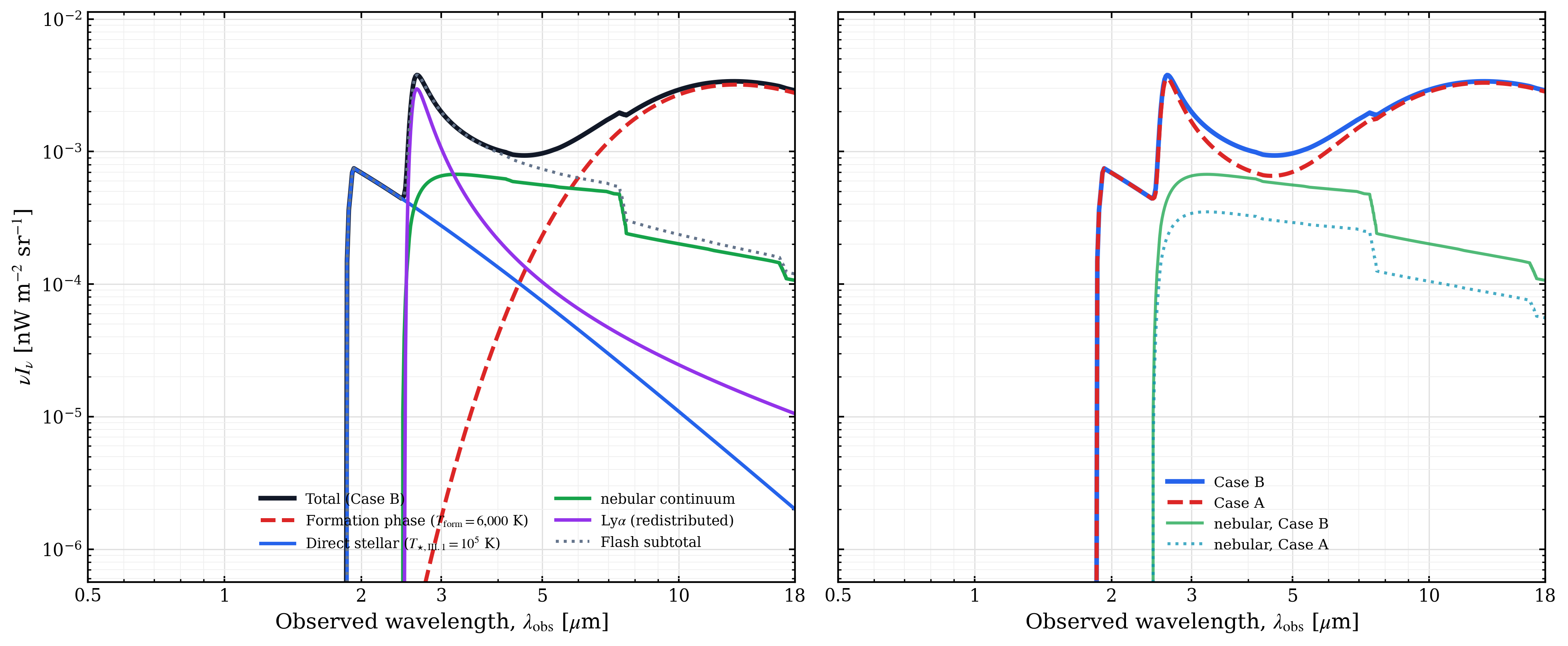}
\caption{Fiducial Pop III.1 supermassive star contribution to the infrared EBL. {\it (a) Left:} Individual components of the Pop III.1 EBL are shown, along with the total, using fiducial parameter values: $m_{\rm \star,III.1} = 10^5 M_{\odot}$; $n_{\rm III.1} = 0.2\: {\rm cMpc}^{-3}$; $t_{\rm form} = 30\:$Myr with $T_{\rm form}=6,000\:$K; and $t_{\rm \star,III.1} = 10\:$Myr with $T_{\rm \star,III.1}=10^5\:$K. {\it (b) Right:} Comparison of the choice of Case A and Case B recombination on the Pop III.1 supermassive star EBL.}
    \label{fig:flashdecomposition}
\end{figure*}

\begin{figure*}[t!]
    \centering
    \includegraphics[width=1 \linewidth]{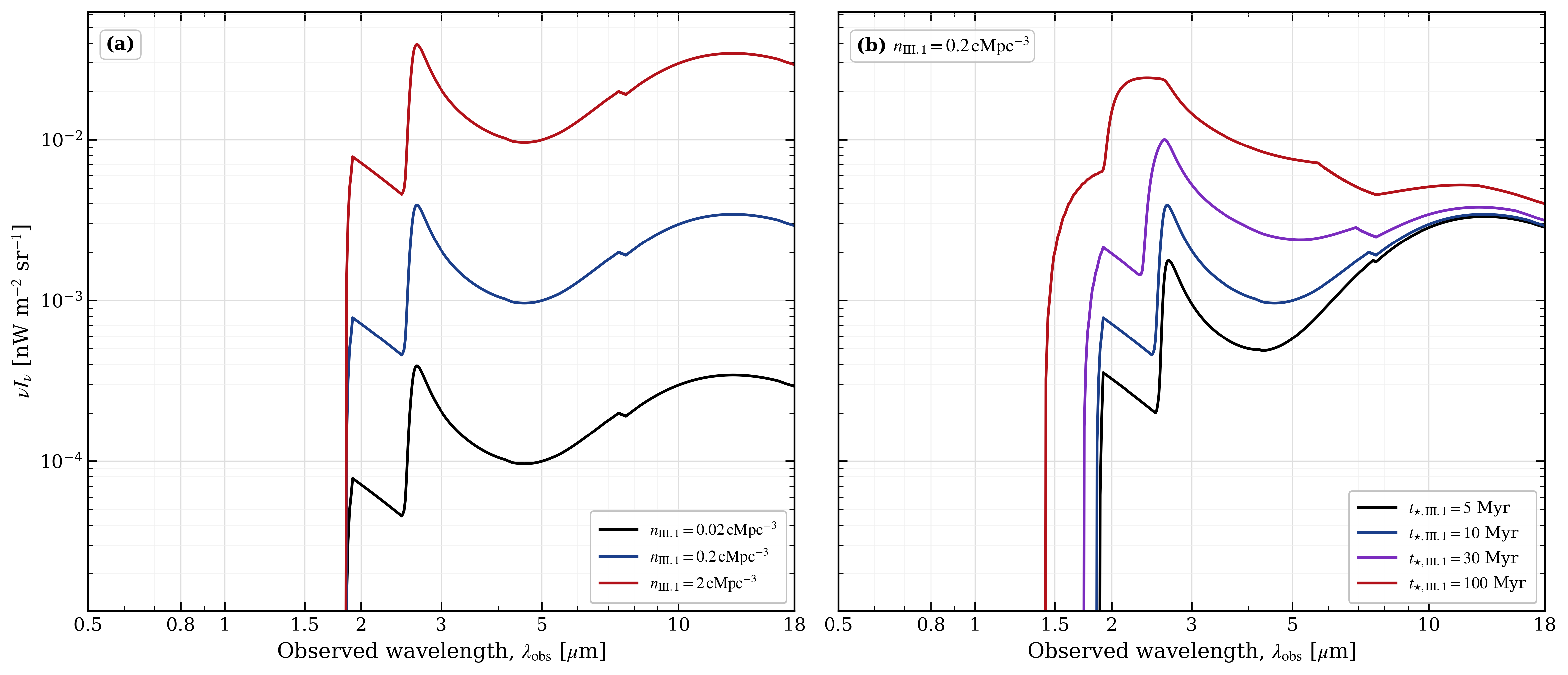}
    \caption{Impact of number density and stellar lifetime on the Pop III.1 supermassive star contribution to the infrared EBL. {\it (a) Left:} Total EBL from fiducial Pop III.1 stellar models, but considering different number densities of $n_{\rm SMBH}=0.02, 0.2$ and $2\:{\rm cMpc}^{-3}$ (bottom to top, as labeled).
    %Observed specific intensity of the Pop III.1 flash component considered for various lifetimes and number densities. We assume the standard parameters $m_{\star,III.1}=10^5\,M_\odot$, $z_{flash} = 20$. Left: different number densities for a fixed lifetime of \(t_{\star, III.1} =10\,{\rm Myr}\). 
    {\it (b) Right:} Total EBL from fiducial Pop III.1 stellar models, but considering 
    different lifetimes at the fiducial number density of $n_{\rm III.1}=0.2$ ${\rm cMpc}^{-3}$. }

    \label{fig:flashlifetime}
\end{figure*}

%\begin{figure}
    %\centering
%\includegraphics[width=1\linewidth]{largemodel2.png}
    %\caption{Comparison model showing that Pop III.1 sources can exceed \cite{salvaterra2003} level specific intensities if one $m_{\star, III.1}= 10^5\,M_\odot$ source is assigned to each eligible Press--Schechter minihalo above \(10^6\,M_\odot\) and allowed to live \(t_{\star, III.1} =400\,{\rm Myr}\). We assume \(z_{\rm flash}=20\), \(T_{\star, III.1} =10^5\,{\rm K}\), and don't consider the formation phase. The intrinsic stellar blackbody is compared with the processed components from considering only nebular continuum, only redistributed Ly\(\alpha\) emission, and both.}
    %\label{fig:placeholder}
%\end{figure}

\subsection{Pop III.1 Stellar Contribution}

We first examine the near-infrared (NIR) and mid-infrared (MIR) EBL signal produced by the fiducial Pop~III.1 model's stellar components. As described above, the baseline calculation assumes a short-lived population of supermassive, Eddington luminosity Pop III.1 stars with $m_{\rm \star,III.1}=10^5\,M_\odot$, comoving number density $n_{\rm III.1}=0.2\,{\rm cMpc}^{-3}$, formation time of $t_{\rm \star,III.1}=30\:$Myr with associated temperature of $6000\:$K and post-formation lifetime of $t_{\rm \star,III.1}=10\,{\rm Myr}$ with associated temperature of $10^5\:$K. The onset of the post-formation phase, which leads to flash ionization of the Universe, is at redshift $z_{\rm flash}=20$. Figure~\ref{fig:flashdecomposition}a shows the resulting observed specific intensity decomposed into the direct stellar, nebular continuum, and redistributed Ly\(\alpha\) components, together with the contribution from the pre-flash formation phase.

The sharp blue cutoff in the direct stellar component is set by the redshifted Lyman limit. At \(z_{\rm flash}=20\), this corresponds to
\begin{equation}
\lambda_{\rm LL,obs} = 912\,{\rm \AA}\,(1+z_{\rm flash}) \simeq 1.92\,\mu{\rm m}.
\end{equation}
Since the sources remain active for \(t_\star=10\,{\rm Myr}\), the emitting window extends to slightly lower redshift, moving the shortest-wavelength edge to \(\simeq 1.85\,\mu{\rm m}\). Of course, in a more realistic model there will be a spread in the formation redshift of Pop III.1 sources, $\Delta z_{\rm flash}$, which will smooth out this edge over a wavelength range $\sim\Delta z_{\rm flash}/z_{\rm flash}$. Considering the results of \citet{Petkova2026PopIIISeedsIV} we expect $\Delta z_{\rm flash}\sim 2$ and thus the short wavelength cutoff to be spread out in wavelength over a range from $\sim1.7$ to $2.0\:{\rm \mu m}$.
%The short wavelength cutoff is further lowered with increasing Pop III.1 lifetime, thus causing the leftward-shifted, more rounded edges of longer-lifetime models.
%, as shown in the right panel of Figure~\ref{fig:flashlifetime}. 
%Thus, 
Immediately redward of the Lyman limit, the direct component is the escaping non-ionizing blackbody continuum. Blueward of this, the direct ionizing continuum is removed, assuming \(f_{\rm esc}=0\), with these photons absorbed locally and reprocessed into the nebular continuum and Ly\(\alpha\) channels. 

There is some uncertainty on the local density of gas that reprocesses the H-ionizing radiation. In the low density limit relevant for large scale ionization of the IGM, it is appropriate to adopt Case A recombination. On the other hand, if densities are higher, then Case B recombination becomes a more accurate description. Figure~\ref{fig:flashdecomposition}b shows the impact of these limiting cases on the nebular and total EBL spectra. We see that, especially for the total EBL spectra, the difference between Case A and Case B is relatively minor.

\begin{figure}
    \centering
    \includegraphics[width=1\linewidth]{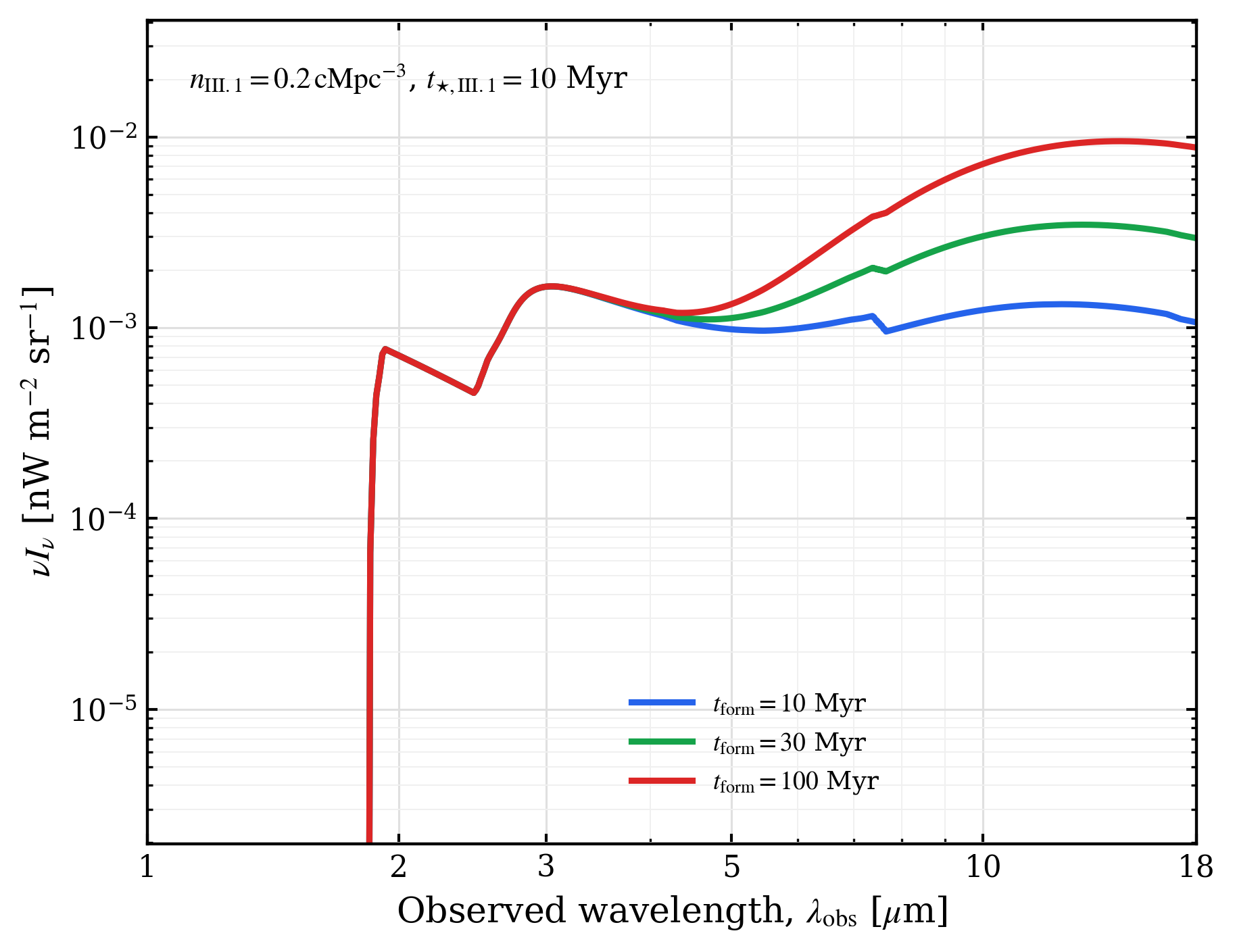}
   \caption{Dependence of Pop III.1 source specific intensity on duration of the formation phase, i.e., for cases with $t_{\rm form} = 10$, $30$ and  $100\:$Myr. Other parameters take their fiducial values, i.e., $m_{\rm \star, III.1}=10^5\,M_\odot$, $z_{\rm flash} = 20$, $t_{\rm \star,III.1} = 10\:$Myr, and $n_{\rm III.1}=0.2\:{\rm cMpc}^{-3}$.}
    \label{fig:growth}
\end{figure}

Between $\lambda \simeq 1.85$ and \(\simeq 2.6 \,\mu{\rm m}\), there is little emission from the nebular continuum and Ly\(\alpha\) channels, leading to this part of the spectrum being dominated by direct stellar emission. At \(\simeq 2.6 \,\mu{\rm m}\), a sharp spike in the specific intensity arises, driven primarily by the  Ly\(\alpha\) channel, whose redshifted peak corresponds to \(\simeq 2.7 \,\mu{\rm m}\) (see Equation~\ref{eq:scattering}). Redward of this peak, the nebular emission then dominates the flash components as the Ly\(\alpha\) contribution rapidly declines away from the redistributed line feature, while the free-bound, free-free, and two-photon continuum terms remain broad in wavelength. This produces a smoother long-wavelength tail in the flash spectrum, in contrast to the narrow Ly\(\alpha\)-dominated spike and the direct blackbody segment immediately redward of the Lyman-limit cutoff.

The location of these spectral features depends directly on the assumed flash redshift. A higher \(z_{\rm flash}\) shifts both the Lyman-limit cutoff and the Ly\(\alpha\)-dominated feature to longer observed wavelengths, while a lower \(z_{\rm flash}\) shifts them blueward. Thus, the observed wavelength of the sharp NIR structure is not only a signature of the radiative process, but also of the redshift at which the flash occurs. In this sense, measuring the relative locations of the Lyman-limit cutoff and the Ly\(\alpha\) feature would provide a direct observational test of when/if the Pop III.1 flash occurred. 

Figure~\ref{fig:flashdecomposition} also shows that the contribution from the protostellar growth phase is dominant at longer wavelengths, i.e., at $\lambda \gtrsim6\:{\rm \mu m}$. This causes the overall EBL spectrum to be relatively flat and reduces the prominence of the Ly\(\alpha\) peak.

As one can see from the parameter survey in Figure~\ref{fig:flashlifetime}, the specific intensity peak from the Pop III.1 flash scales approximately linearly with both lifetime and comoving number density. This allows for significant variation in the specific intensity prediction of the model depending on the assumed values for these parameters, which remain quite uncertain   \citep{Tan2025FlashIonization,2025arXiv250700870N,2025arXiv251000216T}. This approximate linear behavior can be understood from the structure of the emissivity integral. For fixed spectral shape, the amplitude is mainly controlled by the total active stellar mass density integrated over the duration of the source population. Thus, to first order, the Pop III.1 flash amplitude scales with the combination \(n_{\rm III.1}M_{\rm III.1}t_\star\), while the detailed wavelength dependence is set by \(z_{\rm flash}\), the Lyman-limit cutoff, and the redistribution of ionizing photons into nebular and Ly\(\alpha\) emission. This separates the model into an amplitude problem, controlled mostly by abundance, mass and lifetime, and a spectral-shape problem, controlled mostly by redshift and radiative processing. 

%The same linear relationship is observed with mass, but we focus more on the lifetime and number density for our analysis. This is motivated by its reduced uncertainty compared to the other two free parameters; its edge cases differ by less than an order of magnitude. 

Similarly, as seen in Figure~\ref{fig:growth}, the MIR specific intensity scales approximately linearly with the duration of the growth phase. As mentioned, this component dominates at wavelengths $\gtrsim 4\:{\rm \mu m}$, although this is sensitive to the adopted protostellar photospheric temperature of 6,000~K.
%The direct blackbody emission from the growth phase sources becomes relevant after \(\lambda \simeq 4\,\mu{\rm m}\), since this component comes from the same comoving source population but is emitted earlier, during the \(100\,{\rm Myr}\) interval before the flash. It is therefore redshifted more strongly than the flash emission, and because the growth phase has the same comoving number density as the Pop III.1 flash sources, its longer duration allows it to build up a broad contribution at longer observed wavelengths, which can match the flash component's peak at  \(\simeq 2.6 \,\mu{\rm m}\). 
This separation between the hot Pop III.1 flash phase and the cooler protostellar growth phase gives the model a two-part spectral structure: a relatively sharp NIR feature from ionizing radiation and reprocessing during the flash, and a broader MIR tail from the cooler protostellar phase. In this sense, the MIR signal can act as a probe of the growth phase of the Pop III.1 sources. 

\subsection{Pop III.2 Contribution}

\begin{figure}
    \centering
    \includegraphics[width=1\linewidth]{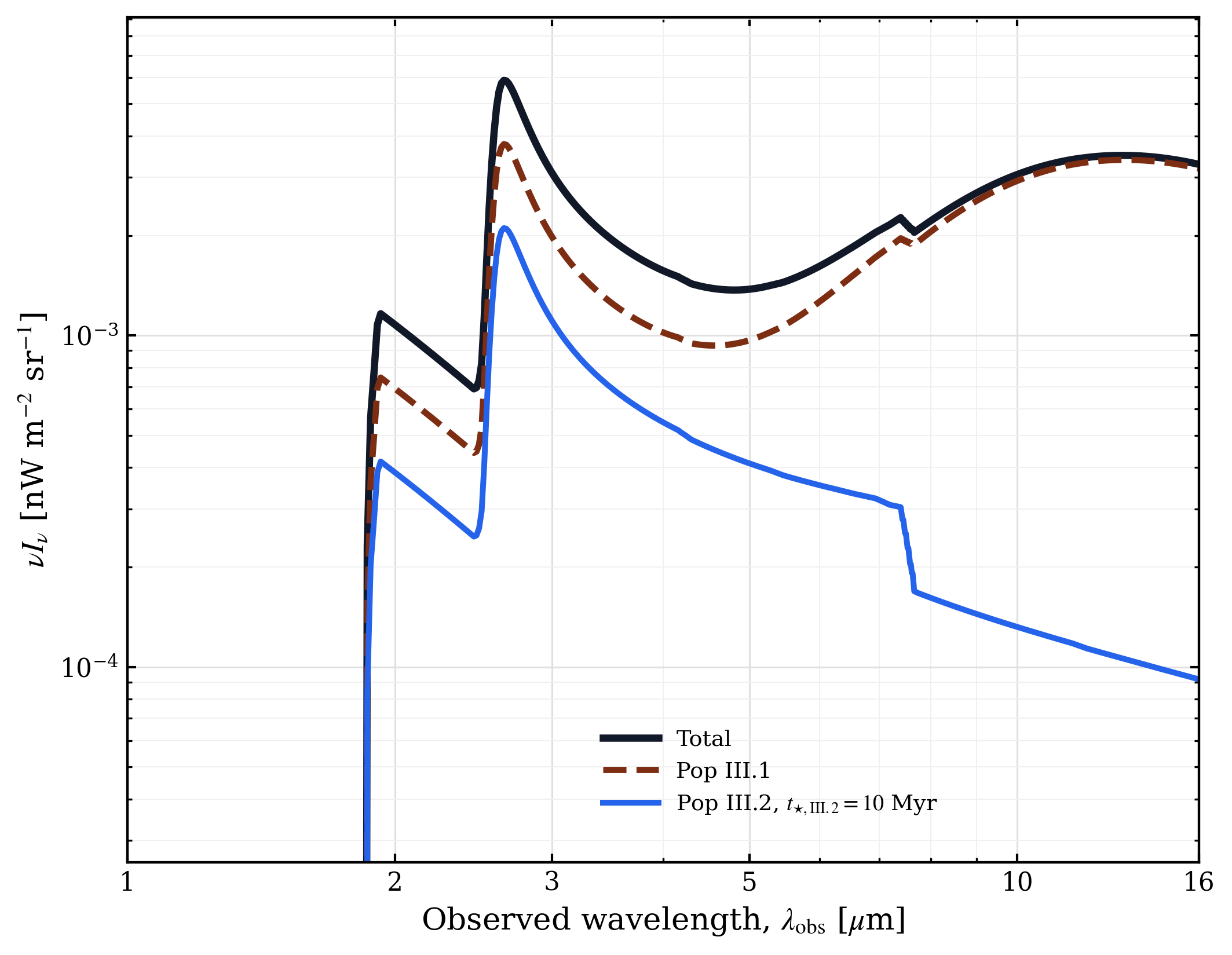}
    \caption{Relative observed specific intensities of the Pop~III.1 and Pop III.2 sources, for the fiducial Pop III.1 model ($t_{\rm form} = 30\:$Myr, $t_{\rm \star,III.1}=10\:$Myr, $n_{\rm III.1}=0.2\,{\rm cMpc}^{-3}$) and a Pop III.2 characteristic mass of $m_{\rm \star,III.2}=100\,M_\odot$ and lifetime of $t_{\rm \star,III.2}=10\:$Myr. }
    \label{fig:poptotal}
\end{figure}

Figure~\ref{fig:poptotal} shows an example EBL spectrum from a model of Pop III.2 sources, assuming $(m_{\rm \star,III.2}=100\,M_\odot$ and $t_{\rm \star,III.2}=10\:$Myr. We see that this Pop III.2 contribution is quite similar (i.e., about 50\%) to that of the fiducial Pop III.1 intensity in the NIR regime. Note, that we have not included any protostellar phase for the Pop III.2 stars, so their contribution to the longer wavelength emission is relatively small.

\begin{figure}
    \centering
    \includegraphics[width=1\linewidth]{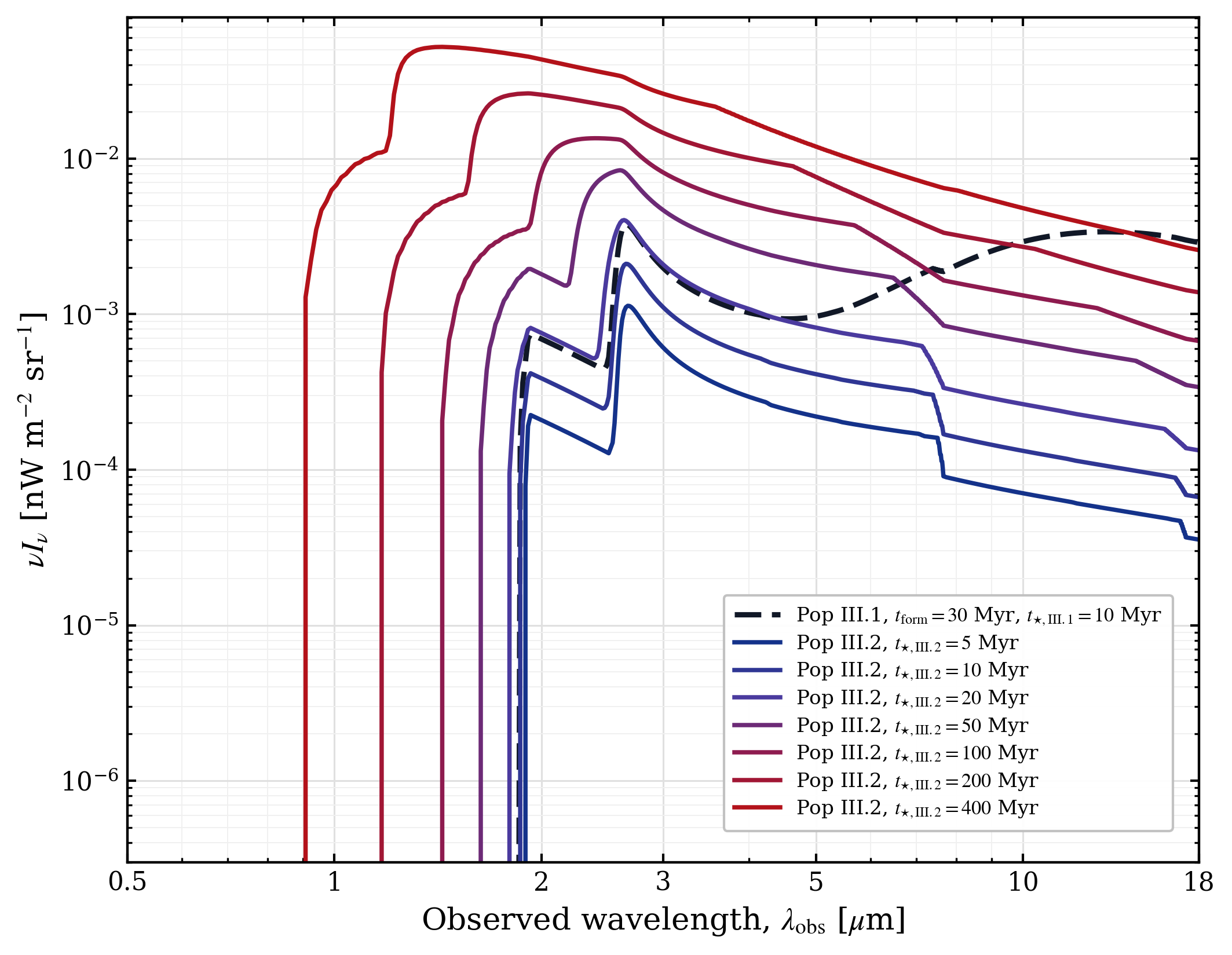}
    \caption{Observed intensity for a parameter sweep of Pop~III.2 lifetimes from 5 to 400~Myr, as labeled. Other parameters of the modeling take their fiducial values, as in Fig.~\ref{fig:poptotal}.
    %    , for $z_{flash} = 20$ and $m_{\star, III.2} = 100 $ $M_{\odot}$. A Pop III.1 source with the assumptions $t_{\star, III.1} = 10$ Myr, $t_{form} = 100 $ ${\rm Myr}$ , $m_{\star, III.1} = 10^5 M_{\odot}$, and $n_{III.1} = 0.2 {\rm cMpc}^{-3}$ is included for reference.  
    }
    \label{fig:popiisweep}
\end{figure}

The Pop III.2 parameters are quite uncertain. In particular, it is possible that these halos host multiple generations of stars (not necessarily all of zero metallicity) that have total lifetimes that are significantly longer than that of an individual star. In this case, the more relevant metric is the total star formation efficiency of the gas in the minihalos. Such an approach was adopted by \citet{salvaterra2003} in their modeling of the ``Pop III'' contribution to the EBL. To show the potential impact of such extended star formation, in Figure~\ref{fig:popiisweep} we show various cases with $t_{\rm \star,III.2}$ ranging from 5 to 400~Myr. If a $100\:M_\odot$ Pop III.2 star has an actual lifetime of 3~Myr, then the total stellar mass implied by these models ranges from $\sim170$ to $13,300\:M_\odot$ and the star formation efficiency (relative to an initial total gas mass of $\sim 200,000\:M_\odot$) ranges from $9\times10^{-4}$ to 0.07. We see that in cases with $t_{\rm \star,III.2}\gtrsim50\:$Myr, the Pop III.2 component dominates over that of Pop III.1 in the NIR. The extended duration of these models also brings the peak of the emission below $2\:{\rm \mu m}$.

%Despite their far smaller mass compared to Pop III.1 sources, Pop III.2 stars can be just as significant to the background due to their significantly higher abundances. Depending on the stellar mass and stellar lifetime assumed, Figure~\ref{fig:poptotal} and \ref{fig:popiisweep} show that their contribution can easily surpass the Pop III.1 observed specific intensity. The Pop III.2 observed spectral shape is also almost identical to the Pop III.1 flash component for similar parameter values, as one would expect from the similarity of the two models. 

%The same linear scaling with number density, source mass, and lifetime is present, as the Pop III.1 sources. This only goes for the NIR, however. The growth phase, which dominates the MIR for the Pop III.1 sources, does not exist analogously for Pop III.2 stars. Therefore, the relative strength of the NIR peak and the MIR continuum may help distinguish whether the background is dominated by a short-lived Pop III.1 flash population or by a more numerous population of lower-mass primordial stars.

\begin{figure*}
    \centering
    \includegraphics[width=1\linewidth]{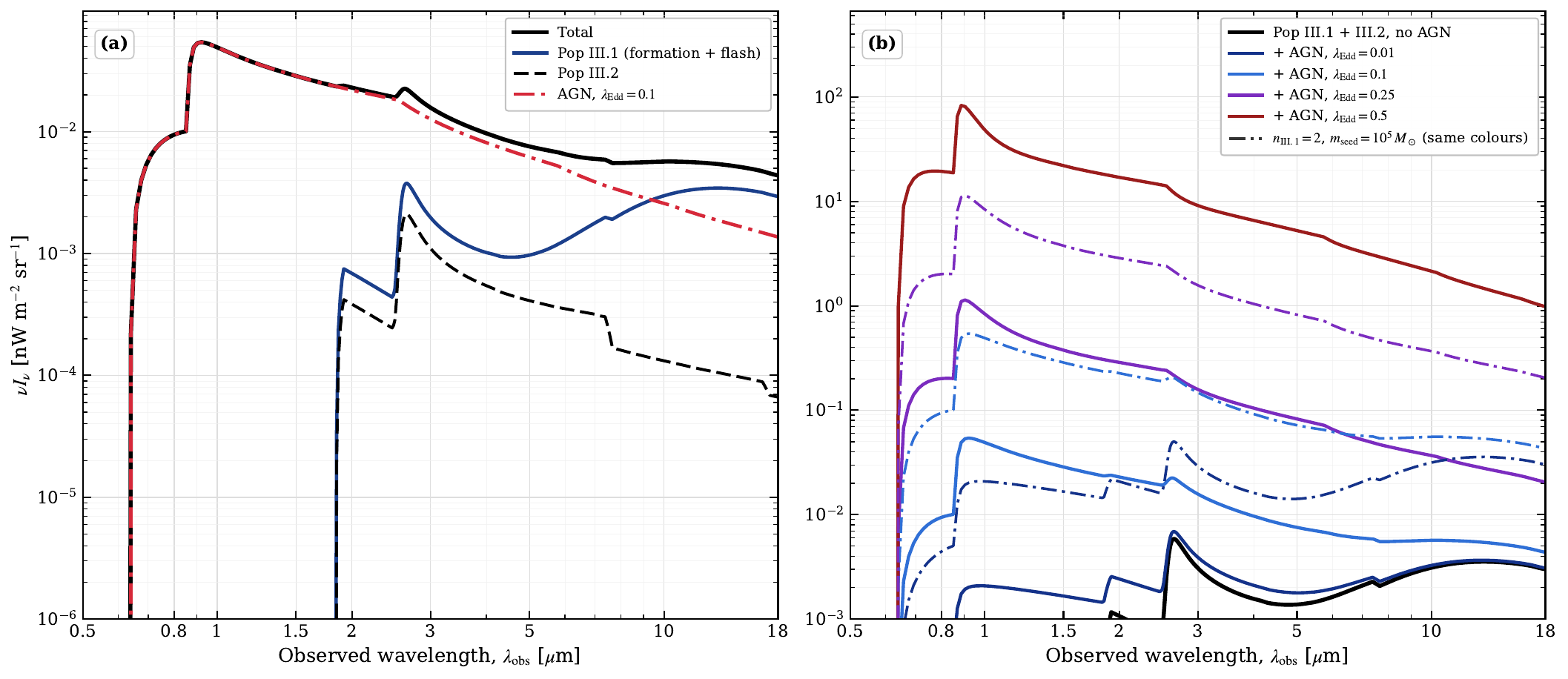}
    \caption{AGN contribution from Pop III.1 remnant SMBH seeds, added to the Pop III.1 and Pop III.2 stellar components. \textit{(a) Left:} Spectral decomposition for the fiducial Eddington ratio $(\lambda_{\text{Edd}}=0.1)$ and number density $n_{\rm III.1} = 0.2\:{\rm cMpc}^{-3}$, showing the Pop III.1 flash component ($t_{\rm \star,III.1}=10\:$Myr), the Pop III.2 component ($t_{\rm \star,III.2}=10\:$Myr), the AGN thin-disk contribution, and their total. \textit{(b) Right:} Total spectrum for Eddington ratios $\lambda_{\text{Edd}}=0.01$ to $1$, compared to the no-AGN baseline.}
    \label{fig:AGN}
\end{figure*}

\subsection{AGN Contribution}

Figure \ref{fig:AGN}a shows an example EBL spectrum from a model including Pop III.1 and Pop III.2 stars, as well as an AGN contribution with $\lambda_{\text{Edd}} = 0.1$ and $z_{\rm end}=6$. We see that such an AGN contribution dominates over the Pop III.1 and III.2 stellar contributions, especially below $\sim 2.5\:{\rm \mu m}$. However, the Ly-$\alpha$ peak of the Pop~III.1 and III.2 stars is potentially distinguishable in the total EBL spectrum. At wavelengths $\gtrsim9\:{\rm \mu m}$, the Pop III.1 protostellar component begins to dominate over the AGN contribution, but with only a gradual change in the spectral index of the EBL.

Figure \ref{fig:AGN}b shows the impact of varying the Eddington ratio, exploring values from 0.01 to 0.5. Over this range, the EBL intensity varies by about 4 dex, with this sensitivity arising because of the exponential growth rate of SMBHs at fixed Eddington ratio (see eq.~\ref{eq:SMBHgrowth}). Figure \ref{fig:AGN}b also shows cases with $n_{\rm III.1}=2.0\:{\rm cMpc}^{-3}$ up to $\lambda_{\rm Edd}=0.25$, which we will see below may be relevant if these sources are to be relevant to explain the reported CIB excess.

We note that the duration of our modeled AGN growth phase is from $t_0=187\:$Myr (i.e., $t_{z=20}+10\:$Myr) to $t_{z=6}=917\:$Myr, i.e., 730~Myr. Thus, the factor by which the average SMBH grows in mass during this phase is $f_{\rm grow}={\rm exp}(730\:{\rm Myr}/t_{\rm acc})\rightarrow 1.2, 4.3, 39, 1500$ for $\lambda_{\rm Edd}=0.01, 0.1, 0.25, 0.5$, respectively (for $\epsilon=0.1$). We expect that the average SMBH mass at $z=6$ is $\lesssim 10^7\:M_\odot$ \citep[][]{2026ApJ...999..175Y}, which limits the growth factor to be $\lesssim 100$ and thus $\lambda_{\rm Edd}\lesssim0.32$ for $\epsilon=0.1$ (and $\lesssim0.71$ for $\epsilon=0.2$).
We also need to consider the SMBH number and mass budget based on present-day constraints. For this, as an example, we adopt the stellar-mass-based black hole mass function of \citet{LiepoldMa2024},
\begin{equation}
  \frac{dn}{d\ln M_{\rm BH}} = \phi \left(\frac{M_{\rm BH}}{M_s}\right)^{\alpha+1}
  \exp\!\left[-\left(\frac{M_{\rm BH}}{M_s}\right)^{\beta}\right],
  \label{eq:bhmf}
\end{equation}
with $\alpha=-1.27$, $\beta=0.45$, $\log\phi=-2.46$ and $\log(M_s/M_\odot)=8.40$.
The mass budget is assessed as the black hole mass density via
\begin{equation}
  \rho_{\rm BH}(>M_{\rm min}) = \int_{\ln M_{\rm min}}^{\infty} M_{\rm BH}\,
  \frac{dn}{d\ln M_{\rm BH}}\, d\ln M_{\rm BH}.
  \label{eq:rhobh}
\end{equation}
%We integrate Eq.~(\ref{eq:rhobh}) directly, %giving $M_\odot$, 
%The integration yields $\rho_{\rm BH}=1.7\times10^{6}\,M_\odot\,{\rm cMpc}^{-3}$ for $M_{\rm min}=10^{5}\,M_\odot$ and $5.7\times10^{5}\,M_\odot\,{\rm cMpc}^{-3}$ for $M_{\rm min}=10^{9}\,M_\odot$, with this latter value being the region of the mass spectrum used to fit by \citet{LiepoldMa2024}. 

Integrating Eqs.~(\ref{eq:bhmf}) and (\ref{eq:rhobh}) down to $M_{\rm min}=10^5$, $10^7$ and $10^9\,M_\odot$ implies comoving number densities of $0.083$, $1.3\times10^{-2}$ and $2.6\times10^{-4}\,{\rm cMpc}^{-3}$, mass densities of $1.7\times10^{6}$, $1.6\times10^{6}$ and $5.7\times10^{5}\,M_\odot\,{\rm cMpc}^{-3}$, and mean masses $\langle M_{\rm BH}\rangle=\rho_{\rm BH}/n=2.1\times10^{7}$, $1.3\times10^{8}$ and $2.2\times10^{9}\,M_\odot$ respectively. 

Thus we see that our models with $n_{\rm III.1}=0.2\:{\rm cMpc}^{-3}$ and $\lambda_{\rm Edd}\sim 0.3$ down to $z=6$ begin to come into tension with the local SMBH budget constraints. However, it should be noted that the  \citet{LiepoldMa2024} mass function is constrained mainly at the high-mass end, i.e., $M\gtrsim10^9\:M_\odot$ and there is significant uncertainty about the properties of the lower-mass ($M\lesssim10^7\:M_\odot$) population \citep[see also][]{Shankar2009,2025ApJ...992..176Z}.

\begin{figure*}
    \centering
    \includegraphics[width=1\linewidth]{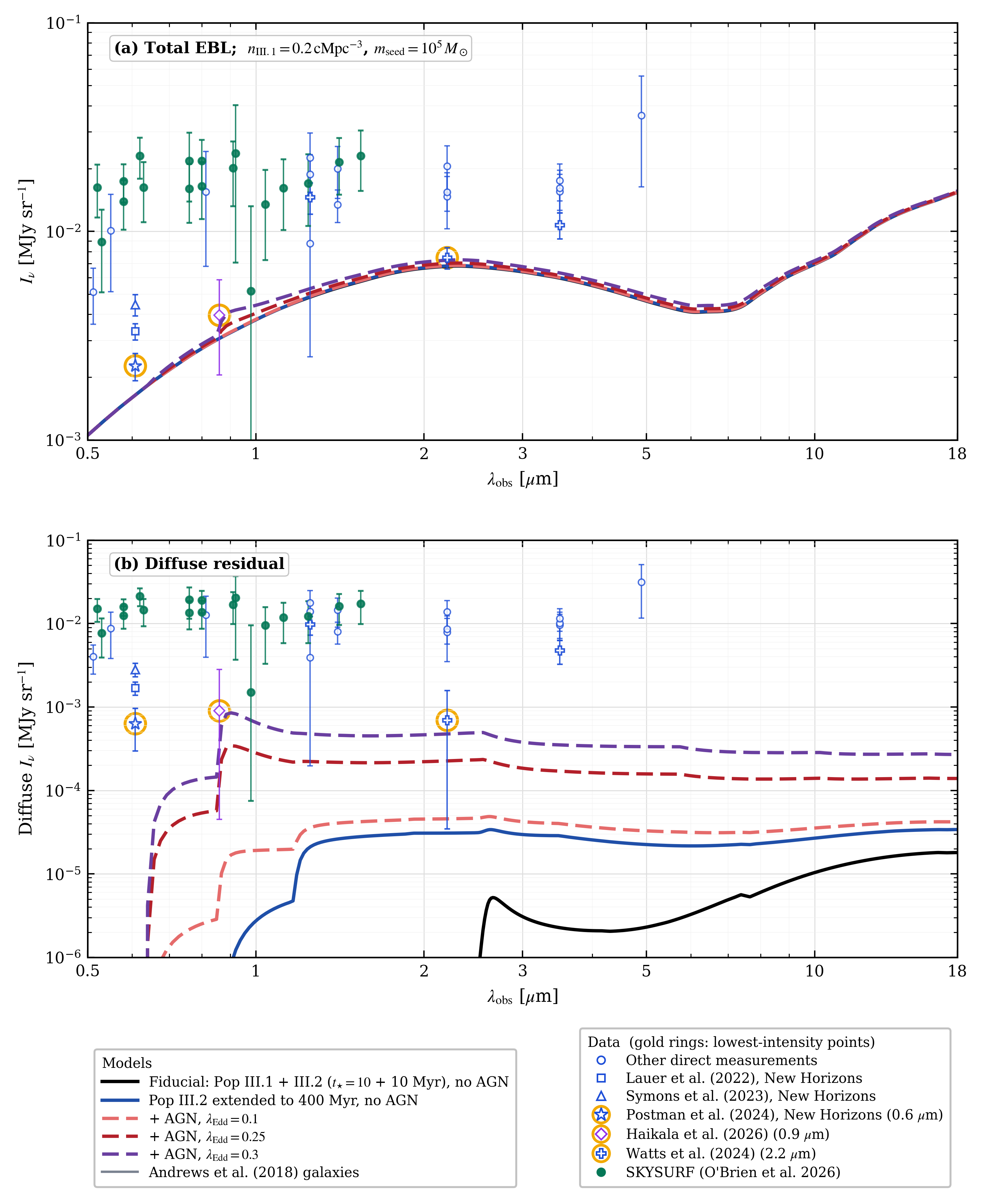}
    \caption{Comparison of the total Pop III.1 contribution to the EBL with observations. We consider the observations presented in \cite{obrien2026skysurf11} and \cite{Haikala2026EBL}, as well as the low-redshift galaxy EBL contribution model in \cite{andrews2018csed}. We include the AGN contribution for the fiducial Pop~III.1+III.2 baseline, a $400 Myr$ lifetime III.2 model, and various AGN models, at \(n_{\rm III.1}=0.2\,{\rm cMpc}^{-3}\). {\it (a)} Total EBL after adding each model to the Andrews galaxy EBL model. {\it (b)} Diffuse excess relative to the Andrews model.}
    \label{fig:mainfig_agn}
\end{figure*}

\begin{figure*}[t!]
    \centering
    \includegraphics[width=1\linewidth]{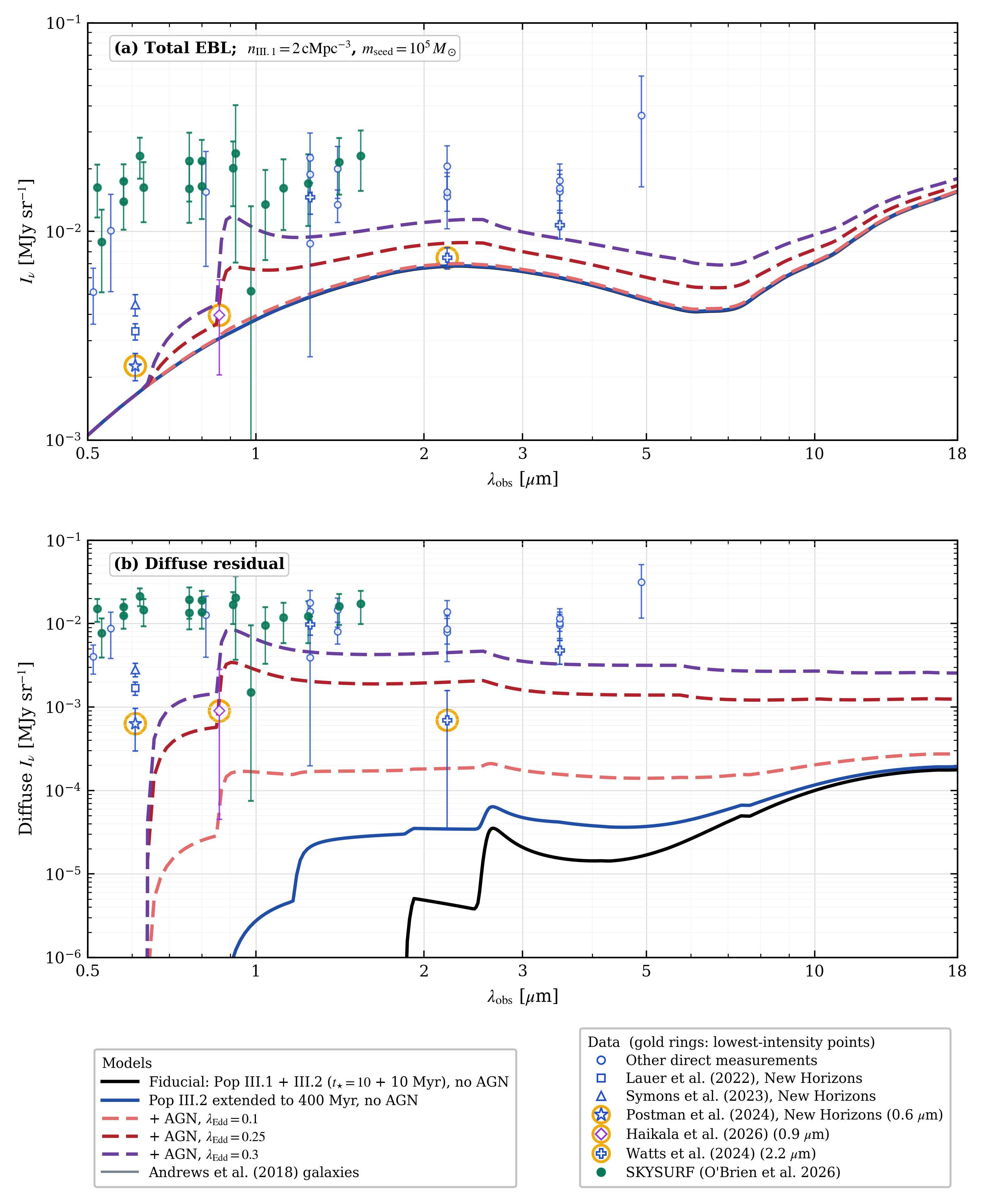}
    \caption{As Figure~\ref{fig:mainfig_agn}, but now with a Pop III.1 comoving number density of \(n_{\rm III.1}=2\,{\rm cMpc}^{-3}\).}
    \label{fig:mainfig}
\end{figure*}

\subsection{Comparison with EBL Observational Constraints}

We compare the contribution of the total Pop~III.1 model, including Pop III.1 and Pop III.2 stellar sources and the resulting accreting SMBH emission from Pop~III.1 remnants, to the SKYSURF and direct EBL measurements presented in \cite{obrien2026skysurf11} \citep[see also][]{
Dube1979,
DwekArendt1998,
Cambresy2001,
ArendtDwek2003,
Bernstein2007,
Levenson2007,
Matsuura2017,
Mattila2017,
Sano2020,
Lauer2022,
Symons2023,
Postman2024,
Watts2024}. In Figures~\ref{fig:mainfig_agn} and \ref{fig:mainfig}, we plot the predicted specific intensities for several plausible models, along with these observational constraints. 

We see that there is significant scatter and uncertainty in the observational data. However, recent measurements of the optical EBL by the New Horizons mission \citep{Lauer2022,Symons2023, Postman2024} provide critical constraints, primarily because they were obtained at heliocentric distances exceeding 40~au, where the otherwise dominant zodiacal light foreground is negligible. Consequently, the primary discrepancy among these measurements stems from their differing treatments of diffuse Galactic light (DGL) contamination. If we adopt the direct EBL intensity of $11.16 \pm 1.65 \, \mathrm{nW \, m^{-2} \, sr^{-1}}$ at $\sim 0.6 \, \mu\mathrm{m}$ (about $2\times10^{-3}\:{\rm MJy\:sr^{-1}}$) from \citet{Postman2024} as the most accurate treatment of DGL, this value imposes the most constraining upper limits on our theoretical models. 
%Specifically, it dictates that any flux contribution from our Pop III.1 flash model cannot exceed this threshold at $0.6 \, \mu\mathrm{m}$.

Similarly, \cite{Haikala2026EBL} have presented a measurement of the EBL at 0.8~microns of about $4\times10^{-3}\:{\rm MJy\:sr^{-1}}$ using the ``dark cloud'' method, i.e., subtracting the spectrum of an opaque dark nebula from an adjacent transparent field.

In the near and mid IR, direct EBL constraints rely almost exclusively on data from the COBE/DIRBE instrument \citep{DwekArendt1998,ArendtDwek2003, Cambresy2001,Levenson2007}. Analyses of these COBE/DIRBE data consistently report an unresolved near-IR EBL excess. For instance, by introducing an additional component to zodiacal light models in an attempt to account for the excess, \citet{Sano2020} still report significant EBL intensities of $21 \pm 4 \, \mathrm{nW \, m^{-2} \, sr^{-1}}$ at $2.2 \, \mu\mathrm{m}$ and $15 \pm 3 \, \mathrm{nW \, m^{-2} \, sr^{-1}}$ at $3.5 \, \mu\mathrm{m}$. More recently, utilizing reprocessed COBE/DIRBE data, \citet{Watts2024} derive lower, yet still robust, values of $10.2 \pm 1.2 \, \mathrm{nW \, m^{-2} \, sr^{-1}}$ ($\sim7\times10^{-3}\:{\rm MJy\:sr^{-1}}$) at $2.2 \, \mu\mathrm{m}$ and $9.2 \pm 1.3 \, \mathrm{nW \, m^{-2} \, sr^{-1}}$ ($\sim10^{-2}\:{\rm MJy\:sr}^{-1}$) at $3.5 \, \mu\mathrm{m}$.

Figure~\ref{fig:mainfig_agn} presents our Pop III.1 EBL models for the case with $n_{\rm III.1}=0.2\:{\rm cMpc}^{-3}$. We see that these models are generally sub-dominant compared to the \citet{andrews2018csed} model for $z<6$ galaxies. Only in the case of relatively high AGN accretion, i.e., with $\lambda_{\rm Edd}=0.3$, does the high-$z$ component begin to be noticeable, with a maximum contribution of about 30\% of the \citet{andrews2018csed} galaxies at $\sim1\:{\rm \mu m}$. Models with larger average Eddington ratios would begin to come into tension with the EBL data. However, as noted in \S\ref{section:AGN}, such models would predict average SMBH masses at $z\sim6$ that are too high compared to observed constraints \citep{2026ApJ...999..175Y}. Thus we see that the fiducial Pop III.1 model, even with relatively high AGN growth rates, is compatible with EBL constraints.

In Figure~\ref{fig:mainfig} we show EBL estimates for example Pop III.1 models with $n_{\rm III.1}=2.0\:{\rm cMpc}^{-1}$. A mild preference for such elevated number densities has been reported by \citet{2026arXiv260619459A} based on patchy kinetic Sunyaev-Zel'dovich constraints of the Pop III.1 flash. We see that these models, with their elevated EBL intensities, are more constrained by the observational data. In particular, models with $\lambda_{\rm Edd}\gtrsim0.25$ begin to exceed the observed constraints in the $\sim0.8$ to $2\:{\rm \mu m}$ regime.

\section{Summary and Discussion} \label{sec:summary}

We have calculated the near- and mid-infrared CIB signature of the Population III.1 supermassive star scenario for SMBH seed formation. Our model includes three physically distinct contributions: (1) the direct stellar, nebular, and Ly$\alpha$-scattered emission from the protostellar growth and near main sequence phases of the $\sim10^5\,M_\odot$ Pop III.1 stars; (2) an analogous but far more numerous population of lower-mass Pop III.2 stars forming in minihalos affected by earlier radiative feedback; and (3) accretion-powered emission from the SMBH remnants left behind once the Pop III.1 stars end their lives.

For our fiducial, physically motivated stellar parameters, the combined Pop III.1 and Pop III.2 signal falls approximately two to four orders of magnitude below existing direct and diffuse NIR/MIR background measurements. The stellar contribution alone is therefore unlikely to explain any potential 
observed CIB excess, unless Pop III source abundances or lifetimes are substantially larger than currently expected. Nevertheless, the stellar spectrum exhibits several characteristic features, including a redshifted Lyman-limit cutoff, a redistributed Ly$\alpha$ peak near $2.6\,\mu{\rm m}$ for $z_{\rm flash}=20$, and a broader MIR contribution from the cooler Pop III.1 protostellar growth phase. These features provide potential observational tests of the Pop III.1 scenario even when the overall signal is subdominant.

However, accretion onto Pop III.1-seeded SMBHs can make a substantial contribution to the overall intensity of the CIB. Since the black holes grow exponentially while accreting at a given fixed Eddington ratio, the predicted AGN contribution depends very strongly on $\lambda_{\rm Edd}$ and on the duration of the accretion phase. For the fiducial number density of SMBHs of $\sim0.2\:{\rm cMpc}^{-3}$, average Eddington ratios of $\lambda_{\rm Edd}\simeq0.3$ lead to EBL intensities that begin to be be constrained by some of the latest observational measurements. However, such accretion rates lead to average SMBH masses that are likely to be too high compared to current estimates at $z\sim 6$. Models with larger number densities, $\gtrsim 1\:{\rm cMpc}^{-3}$ could produce significant fractions of the observed CIB with $\lambda_{\rm Edd}\sim 0.1-0.2$. However, such models, even accounting for mergers, predict the existence of an enhanced abundance of lower-mass SMBHs with $\sim10^6-10^7\:M_\odot$ compared some estimates of local SMBH mass functions \citep[e.g.,][]{Shankar2009,LiepoldMa2024}.

We note that our current modeling does not yet include a full accounting of high redshift galaxy populations beyond the Pop III.1 and Pop III.2 contributions. Such contributions requires more detailed modeling of galaxy formation and evolution, e.g., based on semi-analytic models applied to merger trees \citep[e.g.,][]{Cammelli2025PopIII1SeedsIII} and is deferred to a future study.

We also note that studies of CIB fluctuations that are sensitive to the clustering of the sources are also needed for the Pop III.1 model \citep[see, e.g.,][]{2004ApJ...606..611C,2005Natur.438...45K,2012ApJ...753...63K,2018RvMP...90b5006K,2025ApJ...980L..12K,2025ApJ...994...30F,2026ApJ...999...29K}. In particular, the flash ionization phase is expected to involve R-type HII regions that expand to a characteristic scale of $\sim1\:{\rm cMpc}$, independent of the precise redshift of their emergence \citep{Tan2025FlashIonization}, so this angular scale from emission at $z\sim20$, i.e., $\sim 20\arcsec$, is a feature that is expected to be present in the CIB at $\sim 2.6\:{\rm \mu m}$.

%can account for a significant fraction of the observed excess and near-Eddington accretion can match or exceed it. Thus, while the stellar phases of the Pop III.1 model are unlikely to explain the bulk of the CIB discrepancy by themselves, the associated AGN population may do so if the primordial SMBH seeds undergo sufficiently rapid and sustained accretion. 

Thus, improved measurements of the diffuse NIR and MIR background, both in terms of absolute intensity and its level of fluctuations, especially across the model's predicted spectral features, are needed to further constrain the Pop III.1 SMBH seeding scenario.

%may therefore constrain both the existence of Pop III.1 supermassive stars and the early growth histories of their black-hole remnants.

\section*{Acknowledgments}
 E.T. acknowledges funding from the Virginia Initiative on Cosmic Origins (VICO) through the VICO Undergraduate Summer Research Fellowship. R.O. acknowledges work supported by NASA under award number 80GSFC24M0006. J.C.T. acknowledges support from the Chalmers Initiative on Cosmic Origins (CICO) and the Virginia Institute for Theoretical Astrophysics (VITA), supported by the College and Graduate School of Arts and Sciences at the University of Virginia.
\bibliography{references}

\end{document}